\documentclass[a4paper,twocolumn,amsmath,amssymb,
english,aps,physrev,floatfix,groupedaddress,nofootinbib]{revtex4-2}
\usepackage[T1]{fontenc}
\usepackage[latin1]{inputenc}
\usepackage{amsmath}
\usepackage{babel}
\usepackage{graphics}
\usepackage{amssymb}
\usepackage{mathrsfs}
\usepackage{dcolumn}
\usepackage{epstopdf}
\usepackage{graphicx}
\usepackage{color}

\newcommand{\red}[1]{{\color{black} #1}}


\begin{document}
\title
{
Rare-event properties in a classical stochastic model describing the evolution
of random unitary circuits
}
\author {S. L. A. \surname{de Queiroz}}
\email{sldq@if.ufrj.br}
\affiliation{Instituto de F\'\i sica, Universidade Federal do
Rio de Janeiro, Caixa Postal 68528, 21941-972
Rio de Janeiro RJ, Brazil}

\date{\today}

\begin{abstract}
We investigate the statistics of selected rare events in 
a (1+1)-dimensional (classical) stochastic growth model which describes the evolution of
(quantum) random unitary circuits. In such classical formulation, particles 
are created and/or annihilated at each step of the evolution process, according to 
rules which generally favor a growing cluster size.
We apply a large-deviation approach based on biased Monte Carlo simulations, with suitable
adaptations, to evaluate
\red{(a)} the probability of ending up with a single particle at a specified final time $t_f$,
and \red{(b)} the probability of having particles outside the light cone, defined by a "butterfly
velocity" $v_B$, at $t_f$. Morphological features of single-particle final configurations
are discussed, in connection with whether the location of such particle
is inside or outside the light cone; \red{we find that joint occurrence of both events 
of types (a) and (b) drives significant changes to such features, signalling a
second-order phase transition}. 
\end{abstract}

\maketitle

\section{Introduction}
\label{intro}
In the study of quantum dynamics, one subject of interest is the propagation of quantum
information. Random quantum  circuits are useful in this connection, as they are 
minimally structured models which manage to capture universal properties pertaining 
to entanglement growth.
The spreading of quantum operators under random circuit dynamics can
be measured by the "out- of-time-order correlator" (OTOC)~\cite{lo69}.
It has been established~\cite{nvh18} that both in 1D and in higher spatial dimensions, 
operator spreading and the growth of the OTOC can be mapped to classical stochastic 
growth models. Here we make use of this correspondence, and study the time evolution 
of such classical systems, specializing to 1+1D. 

We are interested in rare events which occur with very small probabilities, also called 
large deviations~\cite{fdh00,t09,dz10}. In order to achieve this we implement a
large-deviation approach to Monte Carlo simulations which involves sampling according 
to a biased distribution~\cite{hm56,b04}. In this way, instead of focusing
on {\em typical} events which, for standard simulations, occur with highest frequency,
the peak of the associated probability distribution is shifted towards the {\em atypical}
ones which are most relevant to our purposes. As described below, our basic implementation
of such ideas closely follows recent work~\cite{ah02,cc01,ah14,sh19}.

This paper is organized as follows. In
Sec.~\ref{sec:model} we present basic features of the model to be studied, as well
as the calculational methods to be employed. Sec.~\ref{sec:results} gives the
results of our calculations.
Section~\ref{sec:conc} is devoted to discussions and conclusions.

\section{Model and Methods}
\label{sec:model}

\begin{figure}
{\centering \resizebox*{2.8in}{!}{\includegraphics*{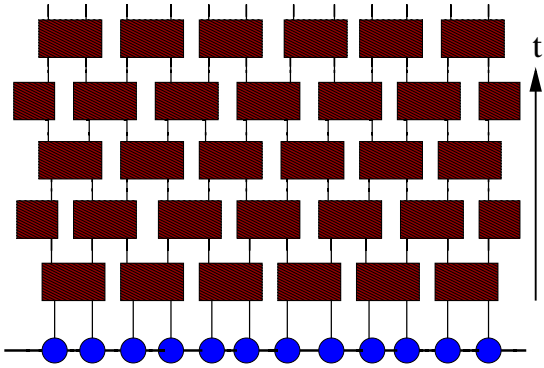}}}
\caption{Lattices sites  in blue. In the RUC problem spins reside on sites,
and the brown rectangles represent independently Haar-random unitaries acting 
on the Hilbert space of two adjacent spins. In the classical growth model the 
rectangles denote bonds between adjacent lattice sites; the latter's occupation 
numbers evolve in time. At each $t$ the corresponding bonds are  updated.  
The staggered placement of rectangles corresponds, 
in the classical model, to the rule that alternating halves of the lattice 
are sequentially updated according to Eq.~(\ref{eq:defprob})
(adapted from Fig.~1 of Ref.~\onlinecite{nvh18}).
}
\label{fig:hru}
\end{figure}

Random unitary circuits (RUC) provide a simplified view of the
quantum process of operator growth. Consider an operator ${\cal O}_0$ (e.g., a
spin) initially localized near the origin; upon Heisenberg time evolution it
will become ${\cal O}_0(t)= U^\dagger(t)\,{\cal O}_0\,U(t)$, acting on many
sites. The "size" of ${\cal O}_0(t)$ is the size of the region in which
${\cal O}_0(t)$ does not commute with some local operator $Y_x$ at position 
$x$~\cite{nvh18}. Assuming that both ${\cal O}_0$
and $Y_x$ are Pauli-like operators, it can be shown that the operator
spread is ruled by the correlator ${\cal C}(x,t)$ given by
\begin{equation}
{\cal C}(x,t)= 1 -{\mathrm Tr}\rho_\infty\,{\cal O}_0(t)Y_x{\cal O}_0(t)Y_x\ ,
\label{eq:otoc}
\end{equation}
where $\rho_\infty$ denotes the infinite-temperature Gibbs state; the second
term, in which the operators are not time-ordered, is the OTOC~\cite{nvh18}.
In such a case, the local operators can be taken to be literally spins localized 
on a lattice. The corresponding pictorial representation of the evolution process 
in one spatial dimension is shown in Fig.\ref{fig:hru}. A spin resides on each blue 
lattice site. Each bond (depicted as a brown rectangle in the Figure) 
represents an independent Haar-random unitary, which acts on the joint Hilbert space of 
two adjacent spins of local Hilbert space dimension $q$ each. Here we shall only 
consider  $q=2$ (Ising spins). Correlations propagate along the lattice via updating 
of bonds. In the quantum system each bond update contributes to ${\cal C}(x,t)$ 
of Eq.~\ref{eq:otoc}.

The classical stochastic formulation in 1+1D which corresponds
to  (quantum) random unitary circuits (RUC) is as follows. 
Referring again to Fig.~\ref{fig:hru},
we start with an infinite chain with sites denoted by integer numbers $\ell$
(the bond between sites $\ell$ and $\ell+1$ is referred to as "bond $\ell$",
for short). 
Initially all sites are empty except for
the "central" one where a single particle resides, so the occupation number
$n_\ell(t=0)=\delta_{\ell,0}$. This corresponds to the operator ${\cal O}_0$
of the quantum description.

The time evolution of this system proceeds via the update of bonds. This means
changing the occupation numbers $\{n_\ell,n_{\ell+1}\}$ of the sites connected to 
bond $\ell$.

An elementary step consists of a simultaneous update of, say, all even bonds 
$2\ell$; the next step is then 
a simultaneous update  of all odd bonds  $2\ell-1$. So all sites are examined once, 
and their occupations possibly changed, at each step. See Fig.~\ref{fig:hru}.

The update rules are as follows:

\noindent\ \red{(R1)}\ $\{n_\ell,n_{\ell+1}\}(t+1)=\{0,0\}$ if  $\{n_\ell,n_{\ell+1}\}(t)=\{0,0\}$.

\noindent\ \red{(R2)}\ if $\{n_\ell,n_{\ell+1}\}(t)$ is any of $\{1,0\}$, $\{0,1\}$,
or $\{1,1\}$ then
\begin{equation}
\{n_\ell,n_{\ell+1}\}(t+1)=\begin{cases}{\{1,0\}\ {\rm with\ probability}\ p}\cr
{\{0,1\}\ {\rm with\ probability}\ p}\cr {\{1,1\}\ {\rm with\ probability}\ 1-2p\ ,}
\end{cases}
\label{eq:defprob}
\end{equation} 
where $p=1/(q^2+1)$~\cite{nvh18}. Here, for Ising spins, $p=1/5$. 
In this case, one may pictorially view the occupation numbers of the classical 
formulation as akin to the individual spin states of its quantum counterpart.
Although particle
annihilation is possible, the above rules favor a cluster growing in size with time.

The ensemble-averaged, time-dependent, length of the cluster, which corresponds to 
the "size" of the OTOC, grows linearly in time with a "butterfly" speed $v_B$.
In 1+1D the exact results of Ref.~\onlinecite{nvh18} show that  
\begin{equation}
v_B=\frac{q^2-1}{q^2+1}\ ,
\label{eq:vb}
\end{equation}
so $v_B=3/5$ in the Ising case. $v_B$ gives the opening of the light cone 
associated with the average operator spread.

Here we investigate the statistics of two species of atypical events 
occurring in the evolution process described above.

\red{(a)} The probability of ending up with a {\em single} particle at a "final" 
time $t_f$.

\red{(b)} The probability and morphology of "super-luminar" configurations, that is,
ones which extend beyond the light cone boundaries defined by $x=\pm v_B t$. 

\red{The physical motivation for the study of the above quantities can be 
understood as follows.
Firstly, given the update rules spelt in Eq.~(\ref{eq:defprob}), and with $p=1/5$,
the minimal final value $N_f\equiv N(t_f) =1$ will only be achieved through some 
very particular
set of concurring, entropy-reducing, fluctuations occurring along the evolution process.
Thus one would expect the intermediate shapes assumed by the cluster at $t < t_f$
to reflect the cumulative contributions of such fluctuations. Conversely, 
realizations with larger $N_f$ would presumably be free of effects of the sort.
Secondly, given that $v_B$ signals the speed of growth of the
ensemble-averaged size of the cluster, excursions outside the light cone 
should result from the accumulation of rare cooperative effects of independent 
fluctuations. Again, the question arises of how the accrual of such effects 
leaves an imprint on the evolving cluster shapes.  
  
Through a quantitative investigation, here we attempt to draw connections
between rare events of types (a) and (b), and the morphological characteristics
exhibited in the evolution of the clusters associated with such events.
In order to do so, we first consider the statistics of $N_f=1$ events in 
Sec.~\ref{sec:nf=1}; general features of super-luminar configurations are given in 
Secs.\ref{sec:slc-1part} for $N_f=1$, and Sec.~\ref{sec:slc-npart} for $N_f > 1$. 
We are then in a position, in Sec.~\ref{sec:slc-nbn}, to     
look at the shape and size of clusters with $N_f=1$,
and check how those properties depend on whether the location of the
single remaining particle is supra- or sub-luminar.

A configuration with $N_f=1$ would correspond, in the quantum RUC model, to the
Heisenberg evolution of an operator in such a way that it ends up having the same 
"size" with which it started (i.e., the length over which it spreads at the end is 
basically the same as the initial one, albeit that it may be shifted from the
origin). Similarly, a super-luminar configuration would correspond to a quantum
evolution process in which the operator correlations spread over atypically
long distances. Since both types of events in the quantum domain would exhibit 
such very specific features, one can expect them to be associated with other 
distinctive traits belonging exclusively to the quantum formulation.     
}

We briefly describe our calculational methods as applied to case \red{(a)}. 
With $P_N(t)$ being
the probability of having a configuration with $N$ particles in total after
$t$ lattice updates, the rules given above allow for $1 \leq N(t) \leq 2t$
with the most probable value $N_{mp}(t) \approx t$. While the region
close to the peak of the distribution is well-described, the 
low-probability tails are much less richly sampled, which tends to compromise 
the accuracy of evaluation of $P_1(t_f)$ if $t_f \gg 1$. 

Following Refs.~\onlinecite{ah02,cc01,ah14,sh19}, we consider a sequence
$Y=(y^{(0)},y^{(1)}, \dots y^{({t_f})})$ of configurations (each consisting
of a collection of occupation numbers $\{n_\ell(t)\}$), where
the transitions $y^{(t)} \to y^{(t+1)}$ follow the update rules; one always
starts with  $\{n_\ell(0)\} = \delta_{\ell 0}$.
$Y$ is called a {\em history}. The quantity of interest for a given history
$Y$ is $N_f(Y) \equiv N(t_f)$, obtained from the associated $y^{({t_f})}$. With 
standard sampling,
each history occurs with its natural probability $R(Y)$.

In the large deviation Monte-Carlo method, sample sequences are generated 
according to a biased distribution $R_\theta(Y)$, in which the probability is
\begin{equation}
 R_\theta(Y) =\frac{1}{Z(\theta)}\,R(Y)\,e^{-N_f(Y)/\theta}\ ,
\label{eq:rtheta}
\end{equation}
where $\theta$ is a fictitious temperature and $Z(\theta)$ is a normalization
factor. For $\theta > 0$ histories with smaller values of $N_f(Y)$ will
be favored; the peak of the biased distribution is thus shifted towards
such configurations with low final density.

One can transform between the biased distribution $P_\theta(N_f)$ and the unbiased 
one $P(N_f)$ which is of ultimate interest by noting that~\cite{sh19}
\begin{eqnarray}
P_\theta(N_f)= \sum_Y R_\theta(Y)\,\delta_{N_f(Y),N_f}=\qquad \nonumber \\
=\frac{e^{-N_f/\theta}}{Z(\theta)}\sum_Y R(Y)\,\delta_{N_f(Y),N_f}
 =\frac{e^{-N_f/\theta}}{Z(\theta)}\,P(N_f)\ .
\label{eq:ptheta}
\end{eqnarray}

One generates the biased distributions $P_{\theta}(N_F)$ according to a standard
Metropolis-Hastings algorithm, through the analysis of a Markov chain of histories
$\{Y(t_{\rm MC})\}$, where the Monte Carlo times $t_{\rm MC}=0, 1, 2 \dots$
denote successive updates $Y(t_{\rm MC}) \to Y(t_{\rm MC}+1)$. For the update,
one generates a trial history $Y_{\rm trial}$, which is accepted  as
the next element $Y(t_{\rm MC}+1)$ of the Markov chain with the Boltzmann-like
probability  
\begin{equation}
A(Y \to Y_{\rm trial}) =\mathrm{min}(1,e^{-\{N_f(Y_{\rm trial})-
N_f \left[Y(t_{\rm MC})\right]\}/\theta})\ .
\label{eq:boltz}
\end{equation}

If $Y_{\rm trial}$ is not accepted, then $Y(t_{\rm MC}+1) =Y(t_{\rm MC})$.
Such transition rules indeed give probabilities as in Eq.~(\ref{eq:rtheta})
and, as is well known, represent the Boltzmann-Gibbs equilibrium distribution 
for a system at temperature $\theta$~\cite{ah02}. 

The lattice update rules of 
Eq.~(\ref{eq:defprob}) make use of (pseudo)-random numbers (RN),
to decide which among the possible outcomes of the examination of each bond 
is chosen. 
In the large deviation approach, the systematics of generating trial 
configurations is as follows~\cite{ah02,cc01,ah14,sh19}.

(i)\ Instead of generating an RN each time one is needed, 
a collection of RNs is computed before running the actual
simulation, i.e., prior to generating a history $Y$, and stored in a
"list", or vector, $\xi$~\cite{sh19}. This list is then consulted in
ascending order, say, each time an RN is required in the build-up of $Y$.  

(ii)\ For a given $t_{\rm MC}$ with its associated
$\xi(t_{\rm MC})$ and $Y(t_{\rm MC})$, a trial vector $\xi_{\rm trial}$  
is generated by randomly replacing a fraction $f$ of the components of
$\xi(t_{\rm MC})$  with freshly-evaluated RNs; this is used in simulating 
a trial history $Y_{\rm trial}$. 

(iii)\ If the latter is accepted according
to Eq.~(\ref{eq:boltz}), then one makes  $\xi(t_{\rm MC}+1)=\xi_{\rm trial}$.
Otherwise a new $\xi_{\rm trial}$ is generated, starting again
from $\xi(t_{\rm MC})$, with the respective $Y_{\rm trial}$ being tested
via Eq.~(\ref{eq:boltz}). 

As a rule of thumb, it is customary to choose $f$ in step (ii) above in such
a way that approximately half the trial histories are accepted~\cite{sh19}.

In this way one probabilistically guides the path
followed by the successive $Y(t_{\rm MC})$, towards a region of configuration 
space where  realizations of the evolution with the desired features are more 
frequent (relative to the unbiased case, which corresponds to $\theta \to \infty$).

By choosing a number of suitable values of $\theta$ one can generate a set of biased 
distributions, each of which is smooth for a specific range of the variable
of interest, in such a way that pairs of said ranges overlap each other to some
(not necessarily very broad) extent. One then uses Eq.~(\ref{eq:ptheta}), and
its straightforward generalization for pairs of finite $\theta$ values, to fit the
respective $Z(\theta)$ by requiring each pair of adjacent distributions to
match one another within the corresponding overlap range~\cite{ah02,ah14,sh19}. 
Starting from the unbiased distribution restricted to the range where it is most
accurate, one then fully reconstructs its full  extent by joining smoothly-shaped
pieces. 

We illustrate the methods just described  by  checking that the distributions studied
here behave in line with the large deviation principle~\cite{fdh00,t09,dz10}.
We do this by evaluating the size-dependent rate function~\cite{sh19,ah11}.
It can be seen that the size-like parameter in our problem is the final time
$t_f$~\cite{nvh18}. It follows from Eq.~(\ref{eq:defprob}) that the maximum 
number of particles at time $t_f$ equals $2\,t_f$. So we take
the density-like quantity  $N(t_f)/2\,t_f$
as the scaled variable to be used when comparing the $P(N(t_f))$
for different final times $t_f$.
Accordingly, we  define the rate function $\Phi (N(t_f))$ as:
\begin{equation}
\Phi (N(t_f)) = -\frac{1}{2\,t_f}\,\ln P(N(t_f))\ .
\label{eq:rate_f}
\end{equation}
According to the large-deviation principle the size dependence of the rate function
should vanish as the system "size" increases~\cite{sh19,ah11}.

\begin{figure}
{\centering \resizebox*{3.3in}{!}{\includegraphics*{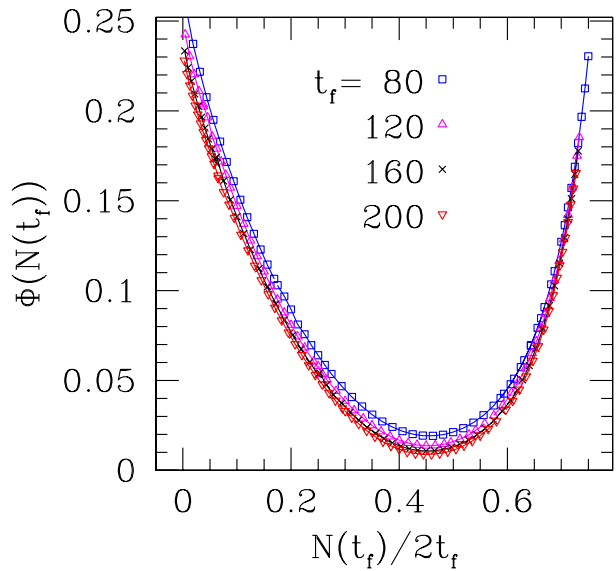}}}
\caption{The rate function $\Phi(N(t_f))$ defined in Eq.~(\ref{eq:rate_f})
for several values of $t_f$. The distributions $P(N(t_f))$ were evaluated
with $10^8$ samples each. In all cases biased distributions, generally 
with $\theta=\pm 1$ and $2$, were used additionally to the unbiased ones (see text).
}
\label{fig:rate_func}
\end{figure}

Figure~\ref{fig:rate_func} shows that our numerical data do behave in the expected 
way, \red{with curves for various $t_f$ becoming closer to one another as 
$t_f$ increases}.
For evaluation of the $P(N(t_f))$ we used biased distributions with both
positive and negative fictitious temperatures $\theta$, corresponding respectively
to the left and right tails of the full curves shown. We generally took 
$\theta=\pm 1$ and $2$, plus $\theta=0.8$ for $t_f=120$ and $1.5$ for $t_f=80$, 
to complement data for the left tail which is of particular interest for
single-particle final configurations (see Sec.~\ref{sec:nf=1} below).

Further comments on details of the implementation of the approach outlined
above, especially those regarding specific aspects of the processes 
studied here, are given in Sec.~\ref{sec:results}.

\section{Results}
\label{sec:results}

\subsection{Introduction}
\label{sec:res-intro}

Examination of the rules given in 
Eq.~(\ref{eq:defprob}) shows that, for a given history, the number and places of
occurrence of non-trivial bond updates [{\em i.e.} differing from $\{0,0\} \to \{0,0\}$]
are  not {\em a priori} fixed. At time $t$ the location $\ell$ of each such 
update (requiring an RN to decide its outcome) depends on the results of previous updates. 
This is in contrast with the cases 
analysed in Refs.~\onlinecite{ah02,cc01,ah14,sh19}, where the total number of RNs required 
for generating each configuration $y^{(t)}$ of a history is fixed.
In such cases one can view each elementary RN replacement, when building
$\xi_{\rm trial}$ from $\xi(t_{\rm MC})$, as analogous to changing the probability
of a given (local) spin flip 
when simulating a magnetic system~\cite{sh19}. The complementary fact that
a non-replaced RN then corresponds to a certain spin which undergoes the same process, 
at the same "microscopic" time $t$, as in the previous history, is in line with the
idea that replacing a few RNs amounts to a "small" local change to the process
being simulated.  

In the present case,
the replacement of 
a single RN may have effects throughout
the extension of the light cone associated with the bond where it occurs.
Thus the "local spin-flip" analogy invoked in previous cases is not
obviously applicable  here.
Even so, we have found that following rules (i)--(iii) given at the end of
Sec.~\ref{sec:model} can be enough to yield the desired outcome of producing
biased sampling as defined in Eq.~(\ref{eq:rtheta}). Furthermore, we have seen
that replacing 
a fraction $f$ between $1.25\%$ and  $2.5\%$ of the components 
of a vector $\xi$ in simulations
of the RUC model has a similar effect, in terms of the acceptance
rate of trial configurations via Eq.~(\ref{eq:boltz}), to replacing of order $10\%$ of the
corresponding vector for other cases~\cite{ah02,cc01,ah14,sh19}. This is very 
likely related to the nonlocal
effects of replacing RNs in the current model, alluded to above. 

In practice we generated an initial $\xi$ slightly overdimensioned, with $M=3t_f^2/4$
components (corresponding to a matrix with $t_f/2$ lines, $3t_f/2$ columns in 1+1D),
so one could be sure that the end of the list would not be reached
during the build-up of a history $Y$. Each time a history is accepted, we record
the locations $\{\ell,t\}$ where RNs were used. For the following trial history 
we randomly replace a fraction $f$ of the RNs associated with the set of previously
used locations. In doing so we are attempting to realize the concept of a "local"
change in the closest possible way, given the update rules. 

In our simulations of RUC evolution the process always starts 
with the same configuration, namely $n_\ell(t=0)=\delta_{\ell,0}$. With standard 
(unbiased)
sampling, this means that there is no need to wait for "thermalization". This
is in contrast with other problems described via Monte Carlo methods, where
in general one starts with arbitrary initial conditions, and must run the
simulation until such a "microscopic" time $t_R$ when those conditions are 
"forgotten", and the system relaxes into a steady state.

\begin{figure}
{\centering \resizebox*{3.3in}{!}{\includegraphics*{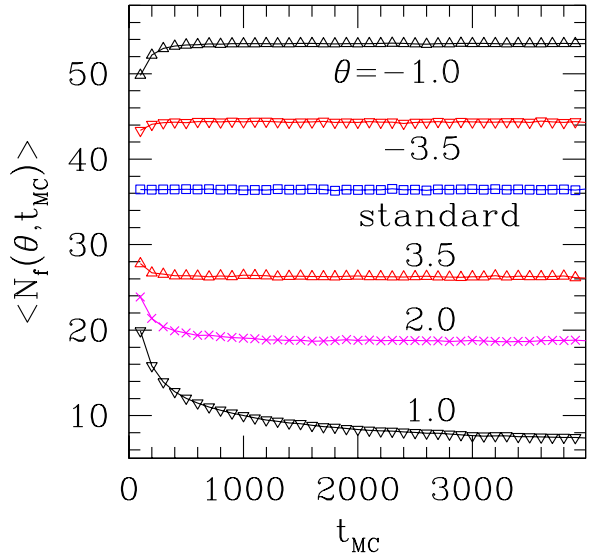}}}
\caption{ $\langle N_f(\theta,t_{\rm MC})\rangle$ are averages, at fixed 
$t_{\rm MC}$, over $10^4$ independent histories of RUC 
evolving up to
$t_f=40$, for various fictitious temperatures $\theta$. The  "standard" label 
corresponds to unbiased sampling, $\theta \to \infty$. For finite $\theta$  
trial vectors $\xi_{\rm trial}$
have a fraction $f=0.025$ of the components of previous $\xi_{\rm MC}$ randomly
replaced (see text). 
}
\label{fig:therm_rel}
\end{figure}

However, for biased sampling as described above, one still has a different
sort of equilibration, related to the fictitious temperature $\theta$, 
as the steps of the Markov chain are correlated via
Eq.~(\ref{eq:boltz}). As stated in Sec.~\ref{sec:model},
rules (i)--(iii) given there ensure that the path followed by the successive
$Y(t_{\rm MC})$ leads the system to the appropriate region of configuration space.
So it takes a finite amount of Monte Carlo time $t_{\rm MC}$ for such
relaxation process to take place.
We illustrate this in Fig.~\ref{fig:therm_rel}, where for each curve we
ran $n_s=10^4$ independent Markov chains up to $t_{\rm MC}=4000$, each chain composed of 
histories starting with a single particle at the origin and evolving to $t_f=40$.
At successive intervals  $\Delta t_{\rm MC}=100$  we evaluated the average 
$\langle N_f (\theta,t_{\rm MC})\rangle$ over the $n_s$ samples. 

As anticipated, the curve corresponding
to unbiased sampling shows no relaxation effects. 

Although our main interest here is the limit of low $N_f$, corresponding
to $\theta >0$, for completeness we looked at thermalization in cases where 
the bias is toward high final densities, enabled via making $\theta <0$.
The curves for $\theta=-1.0$ and $-3.5$ in Fig.~\ref{fig:therm_rel} 
illustrate that the approach to stationarity is qualitatively similar to
that for $\theta >0$, only it is from below. In summary, the bias always 
takes the system's average density away from the unbiased value
given by the blue points in the Figure.

The finite-$\theta$ curves fit reasonably well to
\begin{equation}
\langle N_f(\theta,t_{\rm MC})\rangle =N_f^0(\theta)+ 
a_\theta\,e^{-t_{\rm MC}/\tau_{\rm MC}(\theta)}\ .
\label{eq:relax}
\end{equation}

The adjusted decay times are $\tau_{\rm MC}(\theta)=540(35)$, $204(11)$, $93(10)$
for $\theta=1.0$, $2.0$, and $3.5$ respectively. 
For $\theta=-1.0$ and $-3.5$ one has $\tau_{\rm MC}(\theta)=104(2)$ and $81(10)$.
For $\theta=1.0$ a much improved
fit is found by assuming a superposition of long- and short-lived equilibration processes,
\begin{eqnarray}
\langle N_f(\theta,t_{\rm MC})\rangle =N_f^0(\theta)+ 
a_\theta\,e^{-t_{\rm MC}/\tau^1_{\rm MC}(\theta)}+ \nonumber \\
+b_\theta\,e^{-t_{\rm MC}/\tau^2_{\rm MC}(\theta)}\quad [\,\theta=1.0\,]\ ,\qquad
\label{eq:relax2}
\end{eqnarray}
where $a_\theta$ and $b_\theta$ turn out to be of the same order of magnitude,
$\tau^1_{\rm MC}(\theta)=1076(27)$, $\tau^2_{\rm MC}(\theta)=128(4)$.
We applied  Eq.~(\ref{eq:relax2}) to cases with longer final ("real") 
times $t_f \leq 200$ as well, which will turn out to be relevant
in what follows,  and found that for a range of $\theta \geq 1$ the
longest Monte-Carlo decay time $\tau^1_{\rm MC}(\theta)$ always remains in the
range $(1-2)\times 10^3$. So in our simulations we generally decided to discard the
first $8,000$ MC steps in order to avoid inclusion of non-equilibrated samples.   
Since one needs at the very least some $10^6$ truly equilibrated samples to 
generate suitably smooth probability distributions for application of the 
large-deviation MC approach, 
it is seen that avoiding relaxation effects takes less than $1\%$ of the
total computational effort involved.

\subsection{Statistics of $N_f=1$ configurations}
\label{sec:nf=1}

\begin{figure}
{\centering \resizebox*{3.3in}{!}{\includegraphics*{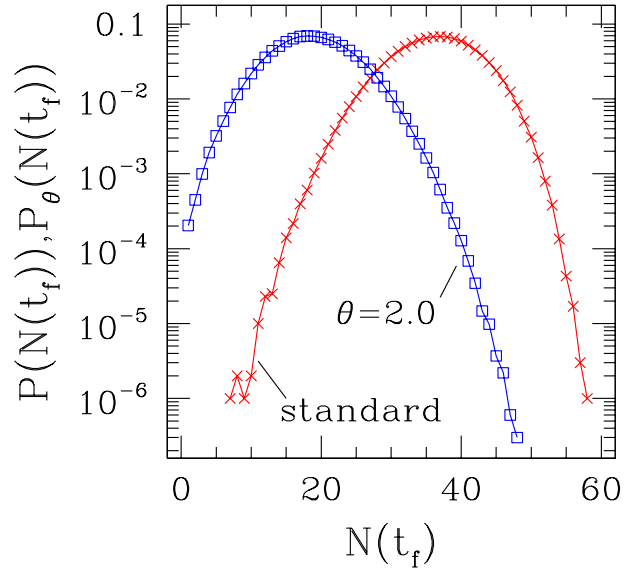}}}
\caption{Probability distributions $P(N(t_f))$, $P_\theta(N(t_f))$, for total number of 
particles after $t_f=40$ lattice updates. The  "standard" label 
denotes unbiased sampling, $\theta \to \infty$. Effective length of Markov chain
corresponds to $10^6$ independent samples for
unbiased case, $10^7$ update attempts (see Sec.~\ref{sec:model}) for $\theta=2.0$.
}
\label{fig:pdn1}
\end{figure}

In Fig.~\ref{fig:pdn1} we show the probability distribution $P_N(t_f)$ for $t_f=40$,
evaluated both with standard (unbiased) sampling and with $\theta=2.0$.
For the latter case, taking into account the results exhibited in 
Fig.~\ref{fig:therm_rel}
we discarded start-up simulation data corresponding to $t_{\rm MC} \leq 2000$.

It is seen that the standard distribution loses accuracy for $N \lesssim 20$.
As already mentioned in Sec.~\ref{sec:model},
in line with the usual procedures of the large-deviation 
approach~\cite{ah02,cc01,ah14,sh19} we search for a range of $N(t_f)$ of common 
support for the biased and unbiased distributions,  where both exhibit smooth
features.  Referring to Fig.~\ref{fig:pdn1}, we find that this corresponds to 
$20 \lesssim N(t_f)\lesssim 35$. Within this range we match both distributions via
Eq.~(\ref{eq:ptheta}), thus extracting an accurate estimate of $Z(\theta)$.

\begin{figure}
{\centering \resizebox*{3.3in}{!}{\includegraphics*{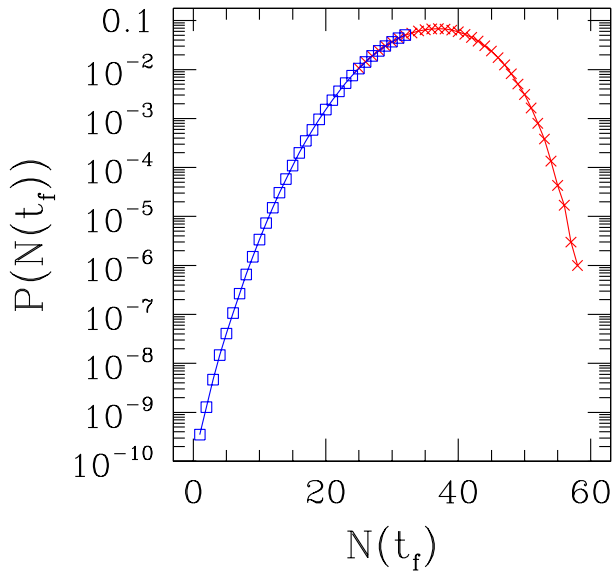}}}
\caption{Result of matching probability distributions $P(N(t_f))$ and 
$P_\theta(N(t_f))$ displayed in Fig.~\ref{fig:pdn1}, following
Eq.~(\ref{eq:ptheta}). Symbol labels are the same as that Figure.
Data are superimposed for $25 \leq N(t_f) \leq 32$, corresponding to the range
actually used for the matching of curves.
}
\label{fig:pdn1_sc}
\end{figure}

Fig.~\ref{fig:pdn1_sc} shows the result of the matching procedure using
$25 \leq N(t_f) \leq 32$, from which one gets $Z^{-1}(\theta)= 958(5)\times 10^3$.
This in turn gives $P_1(t_f)= 3.51(2)\times 10^{-10}$. The largest source of 
systematic uncertainty for the latter quantity seems to be the matching process
itself, rather than the underlying accuracy of [the smooth regions of] the
distributions, which is quite satisfactory when one uses $10^6-10^7$ samples 
as here.

From a Gaussian fit to the $[20-50]$ data range for the standard simulation 
data of Fig.~\ref{fig:pdn1},
one finds the extrapolated estimate $P_1^{\,G}(t_f)=5.9 \times 10^{-10}$. 
This is in line  with
the idea that near the lower bound the distribution must fall more steeply  
than predicted by a Gaussian picture, which assumes the curve to extend
indefinitely to lower values. However, the discrepancy is only by a factor
of order two, suggesting that the Gaussian description works reasonably 
well as a first approximation.

For longer final times $t_f$, one must resort to  values of 
$\theta < 2$ in order to have direct access to $P_1(t_f)$.
The combination of ballistic drift and  diffusive broadening of the front~\cite{ nvh18}
suggests a semiquantitative estimate, based again on a Gaussian
approximation, giving what is expected to be an upper bound for $P_1(t_f)$: 
\begin{equation}
  P_1^{\, G}(t_f) \approx a\,t_f^{-1/2}\exp(-b\,t_f)\ .
\label{eq:gaussian}
\end{equation}   
Using $t_f=40$ data from Fig.~\ref{fig:pdn1} plus results from an unbiased simulation
for $t_f=200$ in Eq.~(\ref{eq:gaussian}), one gets
$a \approx 0.0682$, $b \approx 0.418$.  Thus, for example, 
$P_1^{\, G}(200)=2.37 \times 10^{-39}$.

\begin{figure}
{\centering \resizebox*{3.3in}{!}{\includegraphics*{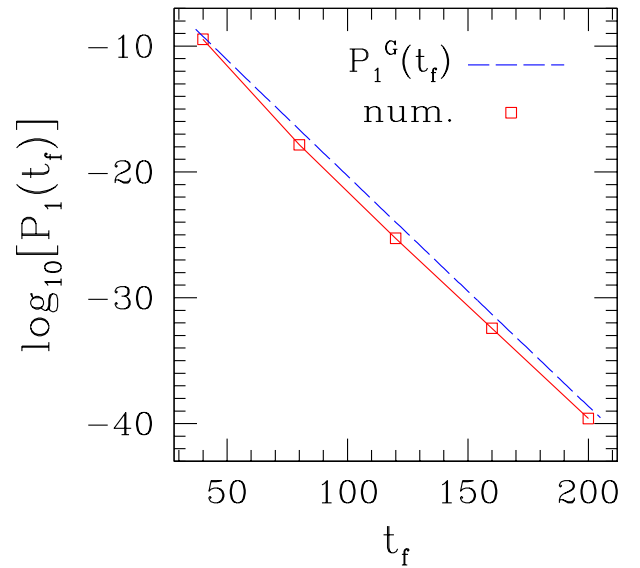}}}
\caption{Red squares are numerical results for probability $P_1(t_f)$
of finding a single remaining particle at $t=t_f$, for $t_f=40$, $80$, $120$, 
$160$, and $200$. Uncertainties are smaller than symbol sizes. The dashed blue 
line is the Gaussian approximation, Eq.~(\ref{eq:gaussian}).
}
\label{fig:p1_nmvar}
\end{figure}

We have  numerically estimated $P_1(t_f)$ for a few  $t_f\leq 200$, which
required use of $\theta$ values as low as $0.8$. The results are exhibited in 
Fig.~\ref{fig:p1_nmvar}, together with the form given in Eq.~(\ref{eq:gaussian}).
It can be seen that the latter indeed acts as a rather stringent upper bound
for the former, exceeding the numerical result by no more than a single
order of magnitude for  $t_f=200$.

\subsection{Super-luminar configurations}
\label{sec:res-slc}

We wish to investigate configurations in which one or more particles
reach a region of space outside the characteristic light cone. Thus,
for evolution up to a final time $t_f$ we search for occupied sites 
with coordinates $x$ such that
\begin{equation}
|x| > v_B\,t_f\ ,
\label{eq:lcone}
\end{equation}
where $v_B=0.6$ is the "butterfly" speed in this case~\cite{nvh18}.

\subsubsection{single-particle configurations}
\label{sec:slc-1part}

Initially, for ease of visualization  we concentrated on the subset of 
configurations with $N_f=1$. In what follows we take $t_f=40$. 
We saw in Sec.~\ref{sec:nf=1} that $P_1(t_f)= 3.51(2)\times 10^{-10}$.
In order to generate richer statistics for the low-$N_f$ end of the
distribution, we used a biased simulation with 
$\theta=1.5$ and $N_s=10^9$ MC steps in all, and selected the resulting
configurations  which ended up with $N_f=1$. On account of the bias we obtained
$P_{\theta=1.5}(N_f=1)=1.9 \times10^{-3}$, which translates to the size of 
the restricted ensemble being $\approx 1.9 \times 10^6$  configurations.   
Note that, since each of the chosen configurations has its intrinsic probability
biased by the {\em same} factor, see Eq.~(\ref{eq:rtheta}), they all have the
same weight in the restricted ensemble.

\begin{figure}
{\centering \resizebox*{3.3in}{!}{\includegraphics*{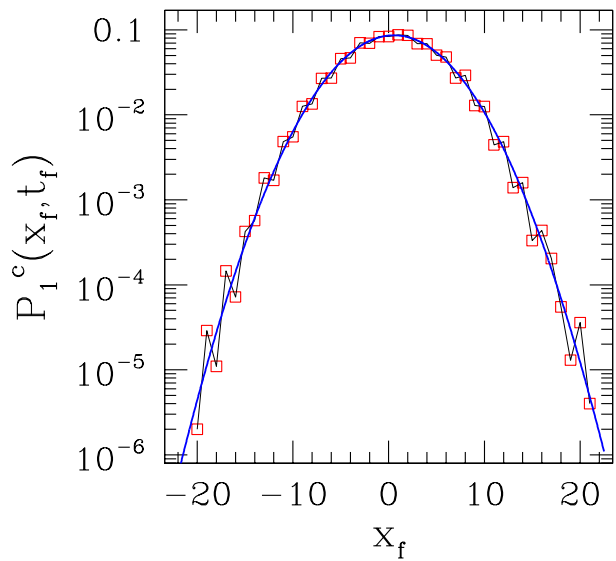}}}
\caption{Conditional probability $P_1^{\,c}(x_f,t_f)$ of having a  particle at 
position $x_f$, given that only a single particle remains,
against position $x_f$ (at $t_f=40$). Restricted ensemble has
$\approx 1.9 \times 10^6$ independent configurations. Blue line
is a Gaussian fit to the data (see text).
}
\label{fig:pdn1_loc}
\end{figure}

Fig.~\ref{fig:pdn1_loc} shows the site-dependent (conditional) probability 
$P_1^{\,c}(x_f,t_f)$ of having the single remaining particle at position $x_f$.
Given the size of the restricted ensemble, we cannot access values
of $P_1^{\,c}(x_f,t_f) \lesssim 10^{-6}$. Consequently the super-luminar
region $|x| > 24$ is out of direct reach. 
The data for pairs of neighboring sites show a tendency towards degeneracy, 
to which we will return later on in Sec.~\ref{sec:slc-npart}. For the moment we
note that, after averaging for such short-range effects,
the available data are remarkably well fitted by a Gaussian
function of width $\sigma=4.62(2)$, shown as a blue line in the Figure. 
Its peak lies within less than half a lattice spacing of the origin. 
Similarly to the case discussed in connection
with Eq.~(\ref{eq:gaussian}), we assume that the Gaussian description remains a
reasonable one away from the region for which it was originally fitted.
Accepting this, one finds that the region contained beyond the boundaries of the 
light cone, {\em i.e.} $P_1^{\,c}(x_f> v_B t_f,t_f)$  corresponds to the area 
below the standard error function beyond 
$|x| \approx 5\sigma$, which is $\approx 10^{-12}$. 
At this point one can return to the original (unbiased) distribution
and estimate, from the data exhibited in Fig.~\ref{fig:pdn1_sc},
 $P_1(|x_f| > v_B t_f) \approx 3.5 \times 10^{-22}$. 

\subsubsection{Configurations with $N_f \geq 1$}
\label{sec:slc-npart}

\begin{figure}
{\centering \resizebox*{3.3in}{!}{\includegraphics*{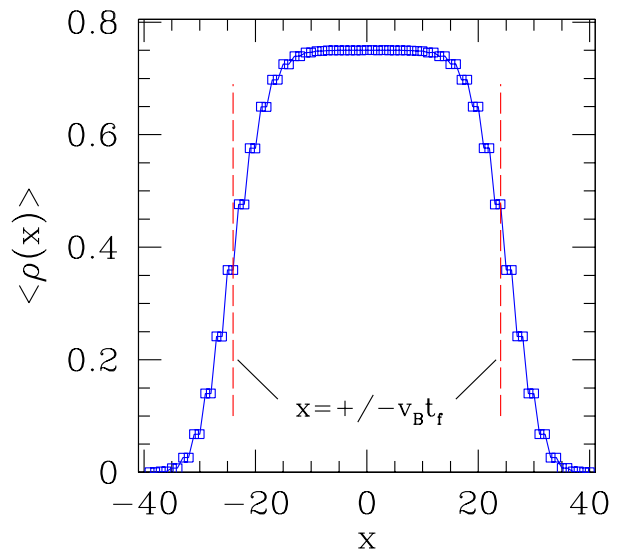}}}
\caption{Ensemble-averaged  local site occupations $\langle \rho(x)\rangle$ 
at $t_f=40$ against position $x$, for $10^7$ independent samples. 
Dashed red lines mark boundaries of light cone at $x=\pm v_B\,t_f=\pm 24$. 
}
\label{fig:loc_dens}
\end{figure}

We now consider the ensemble of all possible final configurations, irrespective
of the total number of particles remaining. We took $t_f=40$ and measured the
site occupation averages $\langle \rho(x)\rangle$ (ensemble-averaged local site 
occupations, with unbiased sampling) at $t_f$ over $10^7$ independent 
samples, with results
shown in Fig.~\ref{fig:loc_dens}. Note a degeneracy of ensemble-averaged densities 
on pairs of neighboring sites. This was already pointed out in connection with
Fig.~\ref{fig:pdn1_loc}, and  is a consequence of the symmetries embedded in 
the evolution rules given through Eq.~(\ref{eq:defprob}).
Such degeneracy is broken (to a quantitatively small degree)
by the introduction of a directional bias, as shown below.

With $N_{\rm out} \equiv N(|x|>v_B\,t_f)$ being the integrated density on sites 
outside the
light cone, and $N_{\rm all}$ the corresponding quantity over the whole lattice,
the relevant fraction is $f_{\rm out} \equiv N_{\rm out}/N_{\rm all}$.

The data exhibited in Fig.~\ref{fig:loc_dens} give $f_{\rm out}=0.07218 \dots$.
Increasing $t_f$ makes the rate of decrease of 
 $\langle \rho(x)\rangle$ around $x=v_B\,t_f$ become steeper. The midway points
where the density becomes half of the plateau value $\langle \rho(0)\rangle$ 
coincide with $|x|=v_B\,t_f$ to a very good extent, as is already the case depicted
in Fig.~\ref{fig:loc_dens}. For $t_f=80$ one gets $f_{\rm out}=0.05373 \dots$.

We wish to generate statistical samples containing many configurations with 
occupied sites outside the light cone.
Defining $N_{\rm out}^L = N(x <-v_B\,t_f)$, $N_{\rm out}^R = N(x >v_B\,t_f)$,
with the corresponding occupation fractions $f_{\rm out}^L$, $f_{\rm out}^R$, 
one possible way to enhance  one of the 
$f_{\rm out}^{L,R}$ is to introduce a {\em directional} bias in the sampling
process. In analogy with the fictitious temperature $\theta$ introduced in 
Eq.~(\ref{eq:rtheta}), which couples to the number of particles $N(t_f)$,  
we introduce a second "temperature" $\tau$ coupled to the position of
the leftmost occupied site at $t_f$, $x_{\rm lm}^f$. 

\begin{equation}
 R_\tau(Y) =\frac{1}{Z_{\rm lm}(\tau)}\,R(Y)\,e^{-x_{\rm lm}^f(Y)/\tau}\ ,
\label{eq:rtau}
\end{equation}
where $Z_{\rm lm}(\tau)$ is once more a suitable normalization constant.
This is the simplest way to incorporate the desired sort of bias. 
Of course different choices could be made for the quantity to be coupled to $\tau$,
such as the center-of-mass location of the full configuration at $t_f$. However,
we shall not investigate such possibilities here.

Again in analogy with the density-biased case described in 
Eqs.~(\ref{eq:rtheta})--({\ref{eq:boltz}), we generate a Markov chain
of histories $Y(t_{\rm MC})$, with updates from $Y(t_{\rm MC})$ to $Y(t_{\rm MC}+1)$
following the generation of trial configurations $Y_{\rm trial}$, for which
the acceptance probability $A_\tau$ is 
\begin{equation}
A_\tau (Y \to Y_{\rm trial}) =\mathrm{min}(1,e^{-\{x_{\rm lm}^f(Y_{\rm trial})-
x_{\rm lm}^f\left[Y(t_{\rm MC})\right]\}/\tau})\ .
\label{eq:boltz2}
\end{equation}

\begin{figure}
{\centering \resizebox*{3.3in}{!}{\includegraphics*{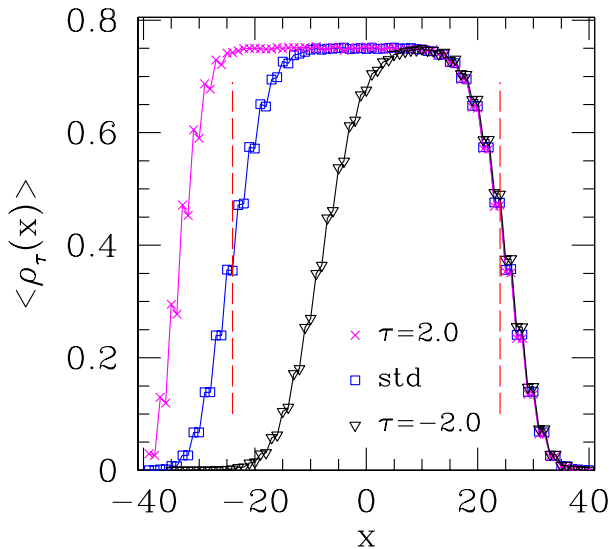}}}
\caption{Ensemble-averaged local site occupations $\langle \rho_\tau(x)\rangle$ 
at $t_f=40$ against position $x$. The "std" label denotes unbiased sampling,
see Fig.~\ref{fig:loc_dens}.
Curves labeled $\tau=2.0$ and $-2.0$  correspond to biased sampling, see 
Eqs.~(\ref{eq:rtau}),~(\ref{eq:boltz2}). $10^7$ independent samples in all cases. 
Dashed red lines mark boundaries of light cone at $x=\pm v_B\,t_f=\pm 24$. 
}
\label{fig:loc_dens2}
\end{figure}

Fig.~\ref{fig:loc_dens2} shows the results of sampling from biased
ensembles, according to Eqs.~(\ref{eq:rtau}) and~(\ref{eq:boltz2}), for
both positive and negative values of $\tau$. For $\tau=2.0$ one gets
$f_{\rm out}^L= 0.1489 \dots$, to be compared to $f_{\rm out}^L= 0.03609 \dots$
for the unbiased case. The biased-sampling scheme thus introduced indeed
affects the overall density outside the light cone, as was the original intention.
For biased cases, see the small breakdown of density degeneracy within pairs of 
neighboring sites, remarked upon earlier.

For a given $\tau$ the transformation between the biased distribution 
$P_\tau(x^f_{lm})$ and the unbiased one goes along the same lines as that given in
Eq.~(\ref{eq:ptheta}) for the case  of $P_\theta(N_f)$, $P(N_f)$. However, 
in order to make a similar  transformation reconstructing the unbiased 
$\langle\rho_{\rm std}(x)\rangle$ of Fig.~\ref{fig:loc_dens2} from
$\langle\rho_\tau(x)\rangle$, further steps are needed.

At a given $x$ the ensemble-averaged density takes contributions
from many biased histories $Y$, each of which has its own value of $x^f_{\rm lm}$
and, associated with it, an exponential weight factor which depends on 
$x^f_{\rm lm}/\tau$.
There is also an overall normalization factor $Z_{\rm lm}(\tau)$, 
see Eq.(~\ref{eq:rtau}). Thus one must keep track of the individual contributions,
for all $x$, of each biased history $Y$ with its own $x^f_{\rm lm}$ and reweight
them with the exponential factor $e^{x^f_{\rm lm}/\tau}$. The normalization
factor can be numerically adjusted at the end.

We proceeded as just described, for the samples used in Fig.~\ref{fig:loc_dens2} 
for $\tau=2$, and reconstructed  $\langle \rho_{\rm std}(x)\rangle$, see the curve
denoted by "std(rec.)" in Fig.~\ref{fig:loc_dens3}.

\begin{figure}
{\centering \resizebox*{3.3in}{!}{\includegraphics*{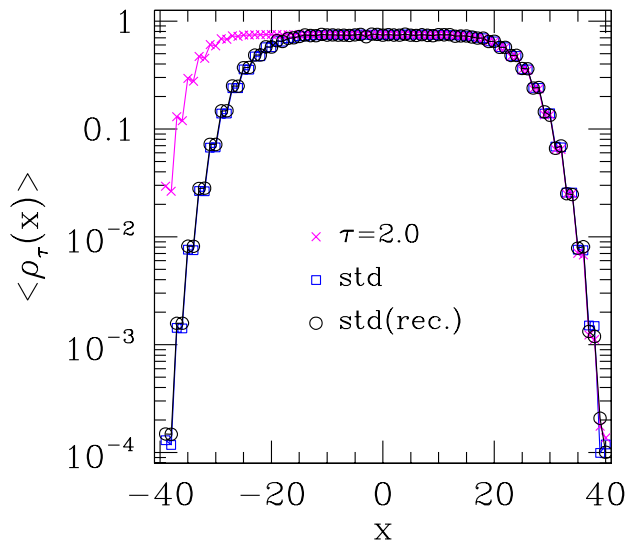}}}
\caption{Ensemble-averaged local site occupations $\langle \rho_\tau(x)\rangle$ 
at $t_f=40$ against position $x$. The "std" label denotes unbiased sampling, 
while the curve labelled $\tau=2.0$ corresponds to  biased 
sampling, see  Fig.~\ref{fig:loc_dens2}; "std(rec.)" denotes 
$\langle \rho_{\rm std}(x)\rangle$
reconstructed from $\langle \rho_\tau(x)\rangle$, $\tau=2.0$ (see text).
Note logarithmic scale for the vertical axis.
}
\label{fig:loc_dens3}
\end{figure}

One sees in Fig.~\ref{fig:loc_dens3} that even close to the cutoff at $x=-40$,
the averaged densities predicted either by the unbiased curve or by the
reconstructed one coincide to within 1-2\%: both are of order $10^{-4}$.
So, $\langle\rho_{\rm std}(x)\rangle$ is already rather accurate for
$|x| \sim t_f$. 

The reasons for this are: (i) there is a hard limit to the range of $x$ beyond 
which particles are not allowed at all, namely $x=\pm t_f$; and (ii) the average 
local density everywhere results
from collecting samples without restrictions over {\em all} accepted histories $Y$, 
each of which may contribute to occupation near the
cutoff regardless of its $N_f$. Compare this, e.g., to $P (N_f)$ for small
$N_f \sim 1$. Owing to the update rules, final configurations with very few
particles are extremely infrequent, so the overwhelming majority of histories do not 
contribute to that in unbiased sampling.

\subsubsection{Configurations with $N_f= 1$: bottlenecks}
\label{sec:slc-nbn}

We now turn to the discussion of morphological features of the connected  
set of successive
configurations assumed by the cluster of occupied sites, as it evolves
from $t=0$ to $t_f$ according to the $1+1$--D RUC growth rules given in 
Eq.~(\ref{eq:defprob}). In what follows, such set will be referred to 
as {\em a configuration}, for short.
Concerning the subset of processes which end with a single particle at
the final time $t_f$, \red{for the Ising case $q=2$ of interest here} 
it has been predicted~\cite{unpub} that a phase
transition takes place \red{in the shape of the associated configurations:} from "fat", 
i.e. a compact shape, for configurations
ending within the light cone \red{($x_f < v_Bt_f$)}, to "thin", i.e. exhibiting a large
density of string-like portions or "bottlenecks",  for super-luminar ones
\red{($x_f > v_Bt_f$)}.

In what follows we adapt the ideas developed in Secs.~\ref{sec:slc-1part}
and~\ref{sec:slc-npart} to the numerical investigation of such predictions. 

One needs to enhance the probability of obtaining the right sort of configuration,
namely one (i) with a single particle left at the end, and (ii) preferably outside 
the light cone. It has been shown respectively in Secs.~\ref{sec:nf=1} 
and~\ref{sec:slc-npart} how either requisite can be fulfilled separately.

We propose the simplest possible {\em ad hoc} generalization, namely
assuming a multiplicative form for the Boltzmann-like factor in
the Metropolis update scheme. That is,
\begin{equation}
 P(N_f,x_{\rm lm}^f) \propto \exp{\left[-(N_f/\theta)-(x_{\rm lm}^f/\tau)\right]}\ ,
\label{eq:theta_tau}
\end{equation}
see Eqs.~(\ref{eq:rtheta}) and~(\ref{eq:rtau}).

To show that Eq.~(\ref{eq:theta_tau}) is appropriate for our purposes in this Section,
first note that we will be  confining ourselves to the restricted ensemble of 
configurations with $N_f=1$; of course the location of that particle will
be  $x_f=x_{\rm lm}^f$. As already argued in connection with Figure~\ref{fig:pdn1_loc},
the factor  $\exp(-1/\theta)$ gives the same bias to 
all  final configurations being considered (and different from that for the ones 
with $N_f>1$). 
Within this subset, those with the single particle further to the left are more
favored; those configurations sharing the same $x_{\rm lm}^f$ are equally biased by the
$\tau -$ dependent factor.

This is convenient, as the {\em relative} probability of occurrence of any two
configurations in the sub-sub-set with (i) $N_f=1$ and (ii) sharing the same 
$x_{\rm lm}^f$ is then, similarly to the case of Figure~\ref{fig:pdn1_loc},
the same as the ratio of their natural (unbiased) probabilities of occurring.

Having this in mind, we ran simulations with the Metropolis probability according to
Eq.~(\ref{eq:theta_tau}) above. 
We  used two sets of $(\theta,\tau)$, namely $(0.6,1.5)$ and $(0.4,1.0)$.
So, particles are pulled to the left of $x=0$ (where the initial particle is).

Fig.~\ref{fig:snap} shows two selected configurations with $t_f=40$ and $N_f=1$, one 
sub-luminar and the other super-luminar. It can be seen that in both cases the qualitative
features predicted in Ref.~\cite{unpub} are unequivocally present: \red{the 
super-luminar (blue) configuration consists mostly of string-like segments, while the
sub-luminar (magenta) one exhibits a broad hull almost everywhere except close to 
$t=0$ and $t_f$ (though with a few internal voids).} 

\begin{figure}
{\centering \resizebox*{3.3in}{!}{\includegraphics*{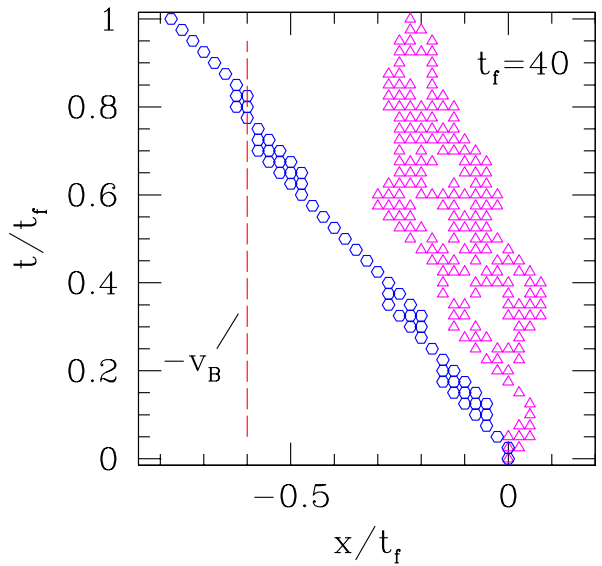}}}
\caption{Spacetime diagrams representing two configurations with $N_f=1$, one 
sub-luminar (magenta triangles) and the other super-luminar (blue hexagons), taken from
a simulation with $\theta=0.6$, $\tau=1.5$, see Eq.~(\ref{eq:theta_tau}), and $t_f=40$.
}
\label{fig:snap}
\end{figure}

We studied the statistics of bottlenecks, i.e., the number $N_{\rm bn}$  of 
constant-$t$ cross-sections of a given $N_f=1$ configuration (other than the 
$t=0$ and $t=t_f$ ones) which have a single particle. 

In Figure~\ref{fig:nbn} the vertical axis gives the
ensemble-averaged number $\langle N_{\rm bn} \rangle$ of bottlenecks according to 
(scaled) final position of single particle $x_f/t_f$, for $t_f=40$, $60$, $80$, and $100$.

For each value of $t_f$ we generated $10^9$ samples in total, for each of the
$(\theta,\tau)$ pairs mentioned above. For all sets of $(t_f,\theta,\tau)$  approximately 
$10^7$ turned out to have $N_f=1$. Because of the low value of the directional-bias
temperature, essentially all $N_f=1$ configurations have $x_f <0$ in this case.
Even so, we have found that those lying outside the light cone are in a clear minority. 
For the data depicted in Figure~\ref{fig:nbn} the largest fraction of 
$N_f=1$ configurations which are super-luminar is $\approx 0.7\%$, for $t_f=40$.
Note that, due to the normalization used, for a given $t_f$ it is appropriate to 
mix $\langle N_{\rm bn} \rangle$ data for both pairs of biasing parameters.
We have seen that results for $(\theta,\tau)=(0.4,1.0)$ behave smoothly for intervals
of $x_f/t_f$ generally farther from the origin than those corresponding to $(0.6,1.5)$, 
though with a reasonably broad intermediate range where such intervals intersect
(the respective data coinciding with one another within small fluctuations).

Figure~\ref{fig:nbn} shows that $\langle N_{\rm bn} \rangle$ increases when $|x_f|$ 
approaches  $v_B t_f$, and (for finite $t_f$) shows a trend to keep increasing beyond 
that; larger values of $t_f$ are associated with steeper growth rates for
$\langle N_{\rm bn} \rangle$.
This is in qualitative agreement with the existing prediction
\red{of a fat-to-thin transition at $x_f=v_B t_f$}~\cite{unpub}.
Defining $\langle N_{\rm bn}\rangle_c$ as the value of $\langle N_{\rm bn}\rangle$
at the (assumedly critical) position $x_f/t_f=-v_B$, the inset of Fig.~\ref{fig:nbn} 
depicts $\langle N_{\rm bn}\rangle_c^{-1}$ against $t_f^{-1}$.
The dashed line is a parabolic fit of the data. Of course, adjusting four points with
three free parameters is only an attempt to probe for underlying trends, the
results of which  should be considered with due caution. 
Estimating an uncertainty of order 
$4\%$ for each data point, the extrapolated intercept of the vertical axis is
$\lim_{t_f^{-1} \to 0} \langle N_{\rm bn}\rangle_c^{-1} = 0.01 \pm 0.02$, with
$\chi^2_{\rm dof}=1.8$. Thus we find indications which are broadly compatible with the 
existence of a 
second-order transition for $\langle N_{\rm bn}\rangle$ at $x_f/t_f=-v_B$, 
for configurations with $N_f=1$.

In summary, our results indicate the applicability of studying the statistics 
of bottlenecks  \red{to provide numerical evidence concerning
the predicted} fat-to-thin transition.

\begin{figure}
{\centering \resizebox*{3.3in}{!}{\includegraphics*{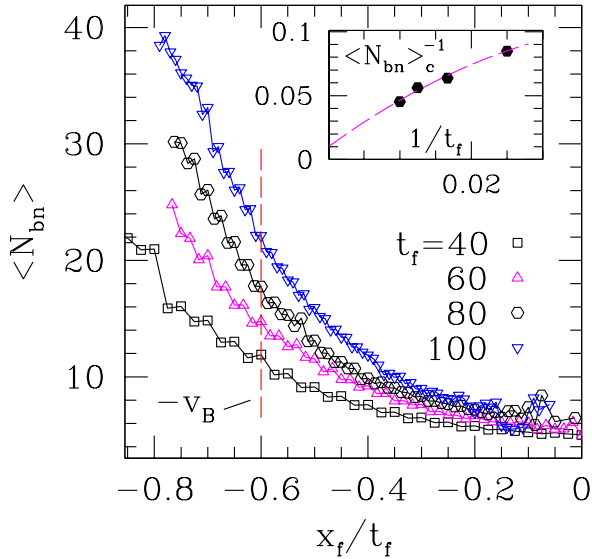}}}
\caption{
Main Figure: the vertical axis gives the
average number $\langle N_{\rm bn} \rangle$ of bottlenecks according to 
(scaled) final position of single particle $x_f/t_f$, for $t_f=40$, $60$, $80$, and $100$.
Averages are taken over the restricted ensemble of configurations with $N_f=1$. 
Data were generated with $(\theta,\tau)=(0.6,1.5)$ and $(0.4,1.0)$, 
see Eq.~(\ref{eq:theta_tau}) and text. Inset: points are $\langle N_{\rm bn}\rangle_c^{-1}$
[$\,$(inverse) $\langle N_{\rm bn}\rangle$ at $x_f/t_f=-v_B\,$]; line is parabolic
fit of data against $1/t_f$ (see text).
}
\label{fig:nbn}
\end{figure}

\section{Discussion and Conclusions}
\label{sec:conc}

We have studied the statistics of selected rare events in RUC evolution processes.
To this end, we have applied a large-deviation approach whose general characteristics 
are well-established~\cite{ah02,cc01,ah14,sh19}. In order to take into account some 
specific features of  RUC evolution, two main adaptations are needed. 

Firstly, the procedure of replacing a single RN of the vector list $\xi$, defined in 
Sec.~\ref{sec:model}, may have effects throughout the extension of the light cone 
associated with the  bond where it occurs. Thus the "local spin-flip" analogy invoked 
in previous applications is not obviously applicable. Even so, we verified that
rules (i)--(iii) given at the end of Sec.~\ref{sec:model} work well here. The main 
difference to other cases in the literature is that, in order to obey the  
rule of thumb of having approximately $50\%$ of the trial histories 
accepted~\cite{sh19}, one needs change only between $1.25\%$ and  $2.5\%$ of the 
components of $\xi$, to be compared to the usual $10\%$ or so used
elsewhere~\cite{ah02,cc01,ah14,sh19}.

Also, in our simulations of RUC evolution the process always starts
with the same configuration, namely $n_\ell(t=0)=\delta_{\ell,0}$. So,
with standard (unbiased) sampling there is no need to wait for "thermalization".
On the other hand, for biased sampling a distinct sort of equilibration takes place.
Fig.~\ref{fig:therm_rel} illustrates that for finite temperature bias $\theta$,
it takes a finite amount of Monte Carlo time $t_{\rm MC}$ for
the path followed by the successive $Y(t_{\rm MC})$ to lead the system to the 
appropriate region of configuration space.

By incorporating the adaptations just recalled, we produced accurate estimates 
of $P_1(t_f)$ as low as $10^{-40}$ for $t_f=200$, which is a suitably large 
value of $t_f$ for our purposes here. Remarkably, we have found the actual 
distribution $P_1(t_f)$ to differ only slightly from a Gaussian shape. 
Fig.~\ref{fig:p1_nmvar} shows that using the latter approximation results 
in a mismatch of at most one order of magnitude out of $40$  
(for $P_1(t_f)$ at $t_f=200$).  

Regarding super-luminar configurations, Figs.~\ref{fig:pdn1_loc}
and~\ref{fig:loc_dens} illustrate that these are in general very unlikely to occur. 
In order to increase the number of realizations with occupied sites outside the light
cone, we introduced in Eq.~(\ref{eq:rtau}) a new temperature-like bias parameter $\tau$ 
associated with a directional bias.

For some quantities of interest, the form Eq.~(\ref{eq:rtau}) is especially convenient
when used together with the idea of analyzing restricted ensembles in which the 
bias factor is the same for all samples considered. Thus their {\em relative} probability 
is unchanged (this latter concept was first used here in connection with the 
unbiased-sampling data of Fig.~\ref{fig:pdn1_loc}). 
One then has the latitude to employ {\em ad hoc} schemes via
Eqs.~(\ref{eq:rtau}) and~(\ref{eq:boltz2}), to evaluate suitable averages
with no need to work back to the original distribution. 

It can be seen, for example, that the ensemble-averaged density distributions shown in 
Fig.~\ref{fig:loc_dens2} tend to coincide at the right end because the
Boltzmann-like factor in Eq.~(\ref{eq:rtau}) was defined as depending only on
the position of the leftmost particle at $t_f$. Considered on its own, such feature
could be deemed as a mere artifact. However, the same definition
comes to one's advantage in Sec.~\ref{sec:slc-nbn}, where one is looking at the subset
of configurations (i) with a single particle at the end and (ii) outside and, say,
on the left of, the light cone. In this particular case, as explained in
Sec.~\ref{sec:slc-nbn}, the choice made ensures that the configurations
used in evaluating bottleneck numbers for Fig.~\ref{fig:nbn} have the correct
relative weights. See Eq.~(\ref{eq:theta_tau}).   

Finally, Figs.~\ref{fig:snap} and~\ref{fig:nbn} lend numerical support to the 
prediction of a fat-to-thin transition for the subset of configurations ending 
with a single particle~\cite{unpub}, depending on where the last particle ends up
relative to the light cone.

\begin{acknowledgments}
The author thanks  the Rudolf Peierls Centre for Theoretical Physics,
Oxford, for hospitality during his visit, and Adam Nahum for many enlightening 
discussions and suggestions. This study was financed in part by Coordena\c c\~ao 
de Aperfei\c coamento de Pessoal de N\'\i vel Superior - Brasil (CAPES) - Finance 
Code 001.
\end{acknowledgments}

\end{document}